\documentclass[slac_one]{revtex4}
\usepackage{graphicx}
\usepackage{fancyhdr}
\begin{document}

\title{Magnetic Scenario for the QCD Fluid at RHIC} %% Paper title goes here


\author{J. Liao , E. Shuryak}
\affiliation{Department of Physics \& Astronomy, SUNY Stony Brook,
NY 11794, USA}

\begin{abstract}
We present new developments of the ``magnetic
scenario''\cite{Liao:2006ry}\cite{Chernodub} for the QCD Fluid
observed at RHIC. The recent lattice data for finite T monopoles
are used to pin down the parameters of the magnetic component of
quark-gluon plasma: in particular we show how the magnetic density
and plasma coupling $\Gamma_M$ change with $T$. We then discuss
one central issue imposed by the heavy ion data: i.e. the
extremely short dissipation length in the QCD fluid around $T_c$.
We show that the Lorentz trapping effect, present only in a plasma
with both electric and magnetic charges and maximal for an equal
50\%-50\% mixture, is the microscopic mechanism leading to the
nearly perfect fluidity. 
\end{abstract}

\maketitle

\thispagestyle{fancy}


\section{E-M DUALITY FOR SQGP} % Section title should be in all capitals.

In very brief term, E-M duality says that a D+1 dim. local field
theory E with D dim. topological excitations M allows a direct
description convenient at weak coupling $e<1$ while has to switch
to a $DUAL$ description at strong coupling $e>1$ in terms of a
local effective field theory which has M as the fundamental
degrees of freedom (D.o.F) and its coupling $g \sim \frac{1}{e} <
1$ (for example \'a l\'a Dirac condition $\alpha_E\cdot \alpha_M
=1 $): for excellent introductions see e.g. \cite{Giacomo_Harvey}.
One good example is the dual superconductivity for color
confinement: QCD vacuum is better considered as a condensate of
monopoles while the perturbative D.o.Fs (i.e. quarks and gluons)
get confined into hadrons \cite{Giacomo_Harvey}.

The main message we've learned from the RHIC program for the last
8 years is that the quark-gluon plasma(QGP) created at RHIC (with
the highest temperature reached about 2$T_c$) is characterized by
fast thermalization and extremely short dissipation length which
means the QGP in 1-2$T_c$ is a very strongly coupled fluid (dubbed
as sQGP). The conclusion is also strengthened by many lattice
calculations. For recent reviews on sQGP, see e.g.
\cite{Shuryak_08}.

Now combining the above two points, one is led to rethink about
what is the most relevant D.o.F for describing the sQGP system.
Indeed there are evidences showing that constituents of the
electric component (quark and gluon quasiparticles) are rather
heavy ($M/T>>1$) already at 1.5$T_c$ and become less and less
important down to $T_c$ and their composite bound states can
survive up to about 2$T_c$ but are heavy
too\cite{Petreczky:2001yp}\cite{bound_states}\cite{Liao:2005pa}.
On the other hand, by borrowing lessons from Seiberg-Witten
theory\cite{SeiWit}, we expect in the deconfined phase there
should be a magnetic component in which monopoles become light,
abundant and weakly coupled when getting very close to the
confining point: in Seiberg-Witten the point is labelled by
specific Higgs while here in QCD the point is $T_c$. The
lightness, abundance and weak coupling should be such as to ensure
the monopoles reach the condensation criteria right on $T_c$ and
enforce confinement. This further leads to two important points:
firstly such a dominant magnetic component in 1-1.5$T_c$ is
necessary and natural for the dual superconductivity in QCD
vacuum; secondly the weakness of magnetic coupling requires via
Dirac condition that the electric coupling must be strong,
explaining theoretically why we have a strongly coupled QGP in
1-2$T_c$. Furthermore there is the ``struggle'' for dominance
between the electric and magnetic sector, fuelled by the opposite
running of electric and magnetic coupling constants: thus there
should be an equilibrium point at about 1.5$T_c$ where E/M
couplings cross each other (both being 1) and the
electric/magnetic components are comparable, while below/above it
the magnetic/electric component wins.

Such a systematic ``magnetic scenario'' for sQGP was first
proposed in \cite{Liao:2006ry}. The existence of a magnetic
component of Yang-Mills plasma and its liquid nature in 1-2$T_c$
was pointed out in \cite{Liao:2006ry} and \cite{Chernodub} from
independent arguments and is confirmed in
\cite{D'Alessandro:2007su}\cite{Liao:2008jg} by analysis of
monopole-antimonopole correlation functions, a traditional and
convincing observable for distinguishing gas/liquid/solid. The
opposite running of E/M couplings was first demonstrated in
\cite{Liao:2008jg} based on data from \cite{D'Alessandro:2007su}.
Lattice evidences for the conjectured equilibrium point around
1.5$T_c$ can be found in \cite{Nakamura} where observables (e.g.
screening mass, spatial string tension, $A^2$ condensate) are
shown to have their E/M dual parts cross around that region. More
recently it has been shown\cite{Liao:2008dk} that the angular
dependence of jet quenching indicates its strong enhancement near
$T_c$ which can be naturally explained only by the ``magnetic
scenario''.

\section{PIN DOWN THE PARAMETERS OF THE MAGNETIC COMPONENT}

\begin{figure*}[t]
\centering
\includegraphics[width=8.cm]{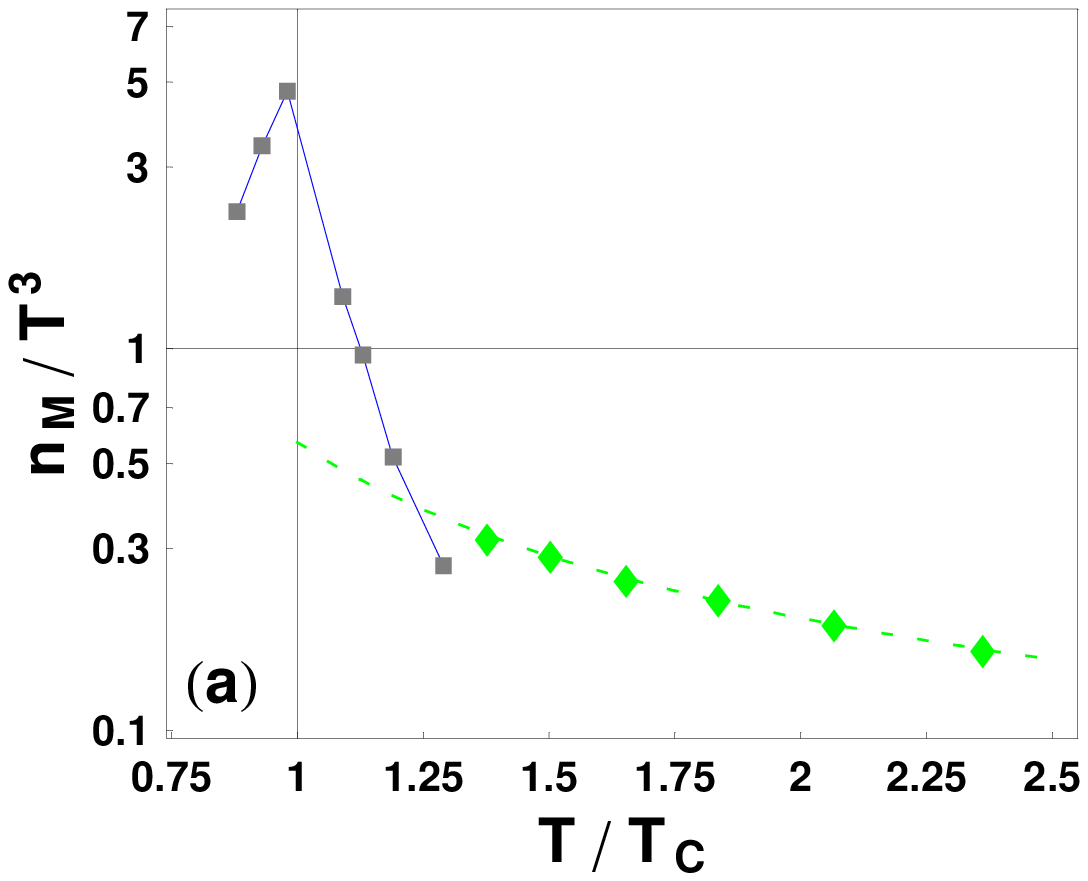} \hspace{0.1in}
\includegraphics[width=6.cm]{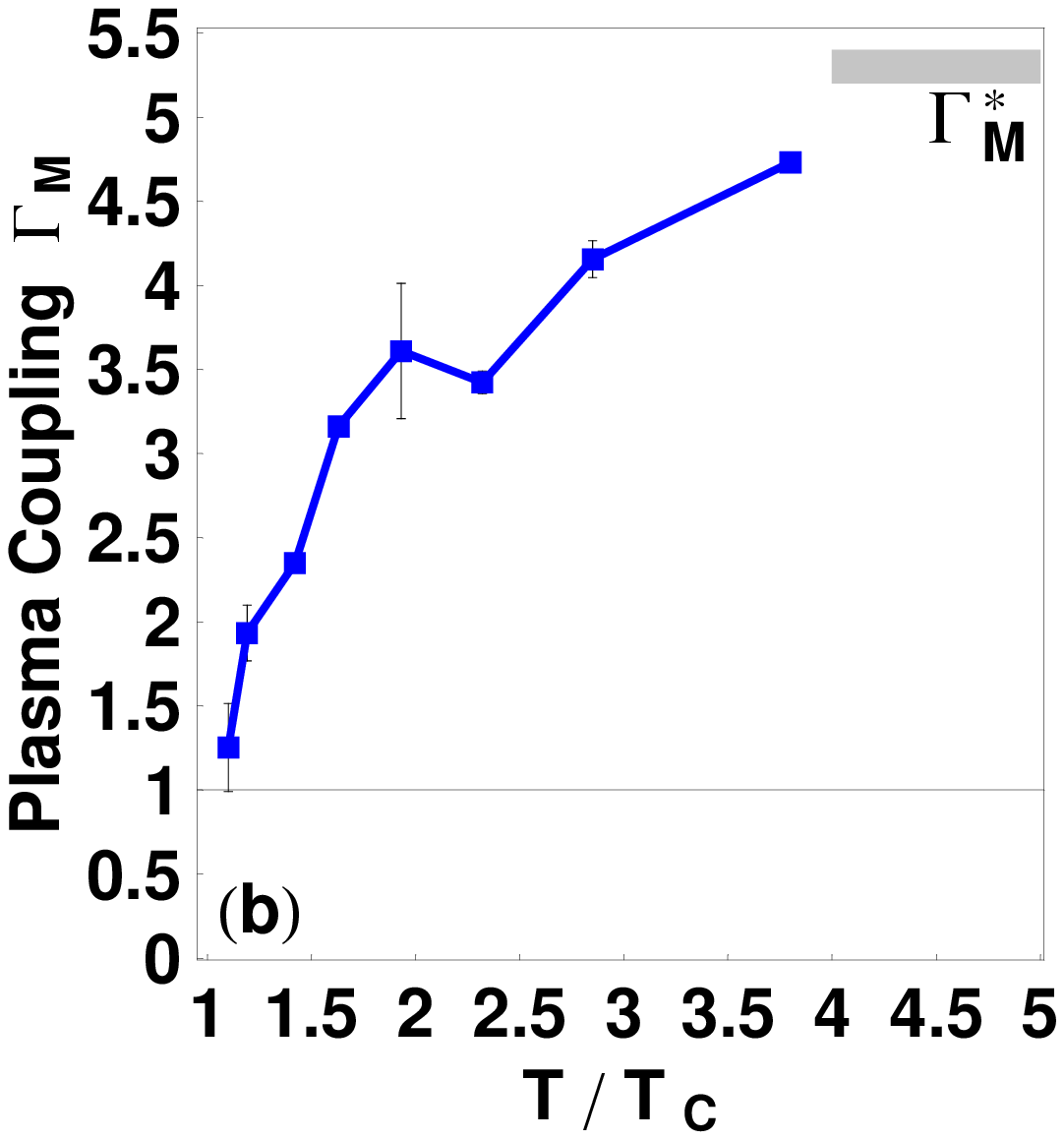}
\caption{(a) Monopole density (on Log scale): the diamonds with
dashed curve are lattice data \protect\cite{D'Alessandro:2007su},
and the boxes with solid curve are from model
calculation\protect\cite{Liao:2008vj}; (b) Magnetic plasma
coupling $\Gamma_M$, the grey band indicates $\Gamma_M^*$ at high
$T$ limit.} \label{parameters}
\end{figure*}

To understand the properties of sQGP, it is crucial to know the
parameters of its magnetic component, i.e. the monopole density,
magnetic coupling, monopole mass, etc. To that end, lattice gauge
theory is a unique and reliable approach. The excellent lattice
results reported in \cite{D'Alessandro:2007su} provided important
information on monopoles' density $n_M(T)$ and their spatial
correlation in a wide region 1.3-4$T_c$. Analysis in
\cite{Liao:2008jg} extracted the magnetic coupling $\alpha_M(T)$
and the plasma coupling $\Gamma_M\equiv \alpha_M (4\pi
n_M/3)^{1/3}/T$ of the magnetic component: $\Gamma$ is also a
well-known observable that can distinguish a gas ($\Gamma<1$),
liquid ($\Gamma\sim 1-10$), glassy matter($\Gamma\sim 10-100$) and
solid ($\Gamma>100$). Another development concerns the strong
linear rise observed in the lattice calculated static $\bar Q Q$
potential energy: the linear part persists and splits from the
free energy linear part (which ceases out when heated to $T_c$) in
0.8-1.3$T_c$ and its slope, defined as an effective tension
$\sigma_V$, peaks at $T_c$ with a value 5 times the vacuum string
tension. This was interpreted in \cite{Liao:2008vj} as formation
of electric flux tube in a magnetic plasma in that T-regime due to
the presence of dense {\it thermal} monopoles. Based on a model
calculation the plasma monopole density is related to $\sigma_V$
and from lattice data for $\sigma_V$ one can infer the monopole
density in 0.8-1.3$T_c$. Some results are summarized in
Fig.\ref{parameters} and below:

\begin{itemize}
\item {\bf Density} [Fig.\ref{parameters}(a)] ---  $n_M/T^3$
decrease at higher $T$ while soars close to $T_c$, as expected. It
is a very densely packed liquid 1-2$T_c$ and especially below
1.5$T_c$, much denser than the relativistic massless ideal boson
gas limit $n/T^3=0.1218$ (this was the original argument in
\cite{Chernodub} for a liquid while we showed in
\cite{Liao:2006ry} the transport properties is in liquid regime).
As comparison the quarks/gluons get denser and denser to high $T$
end.
 \item {\bf Correlation}--- monopole-antimonopole correlations obtained from
 both lattice calculation\cite{D'Alessandro:2007su} and our MD
 simulation\cite{Liao:2008jg}
 show similar shape and peak magnitude that are typical for a liquid, thus providing
 convincing proof of the liquid nature. In particular
 the lattice data showed the correlation grows {\it stronger} at higher
 $T$. Such liquid correlation was also found in recent lattice study\cite{Meyer:2008dt} where no Abelian
 projection was involved.
 \item {\bf Magnetic coupling} --- the extracted magnetic Coulomb
 coupling as a function of $T$ indeed runs in the direction
$opposite$ to the electric one, again as expected, and at high $T$
end it is roughly inverse of the asymptotic freedom formula for
the electric one (see \cite{Liao:2008jg} for details). \item {\bf
Plasma coupling} [Fig.\ref{parameters}(b)] --- down toward $T_c$,
$\Gamma_M$ decreases but remains $>1\,$ i.e. in the good liquid
regime with low viscosity which nicely agrees with empirical
evidences from RHIC experiments. Going to higher $T$ $\Gamma_M$
increases and shows tendency to saturate at a limiting
$\Gamma_M^*$ as we expect from the magnetic scaling valid at very
high $T$ together with the Dirac condition, i.e. $n_M^* \sim
(\alpha_E \, T)^3$ thus $\Gamma_M^* \sim \alpha_M \cdot
(n_M^*)^{1/3}/T \to$ constant $\sim 5$.
\end{itemize}

\section{LORENTZ TRAPPING EFFECT MAKES THE ``PERFECT LIQUID''}

\begin{figure*}[t]
\centering {\includegraphics[width=4.5cm]{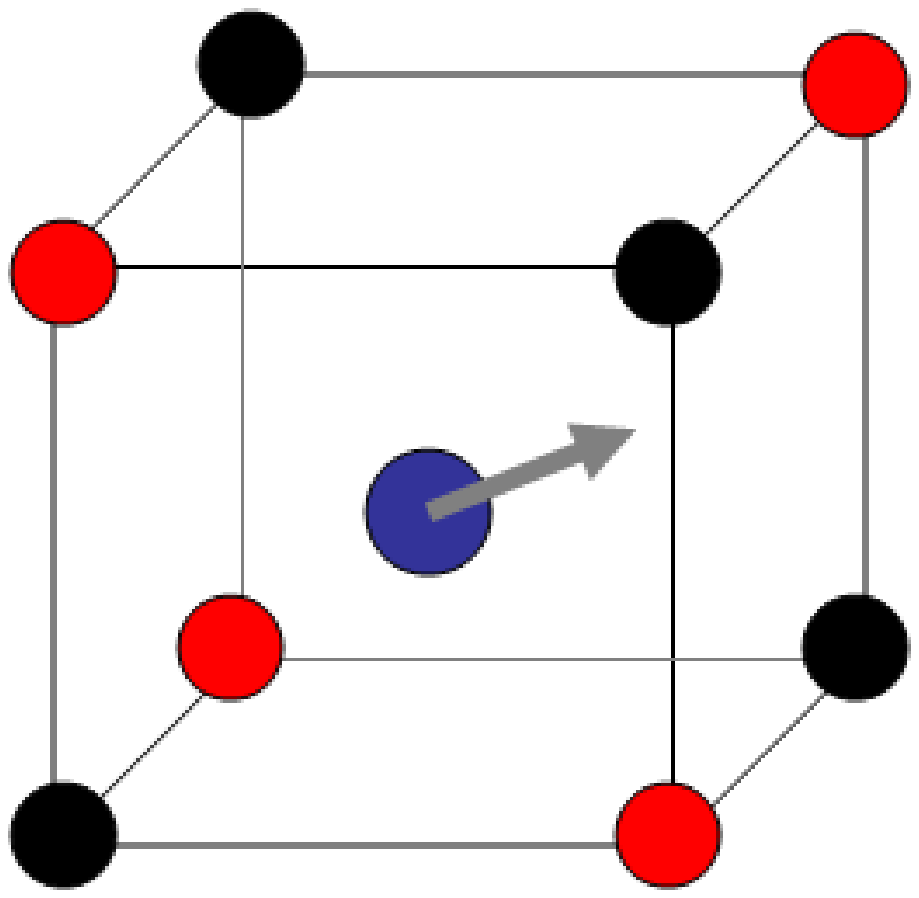}
\hspace{0.2in}
\includegraphics[width=6cm]{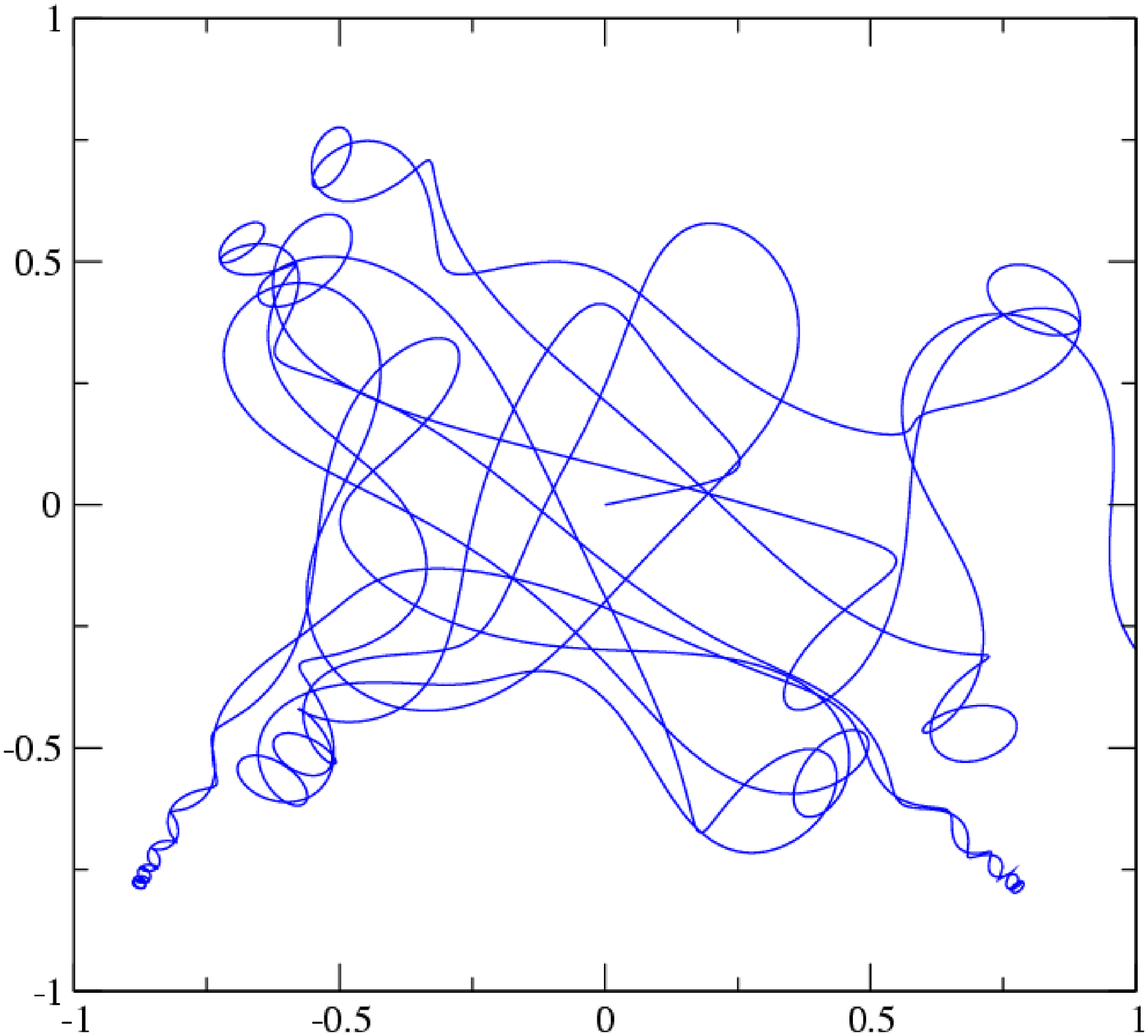}\\}
\centering { \includegraphics[width=6.cm]{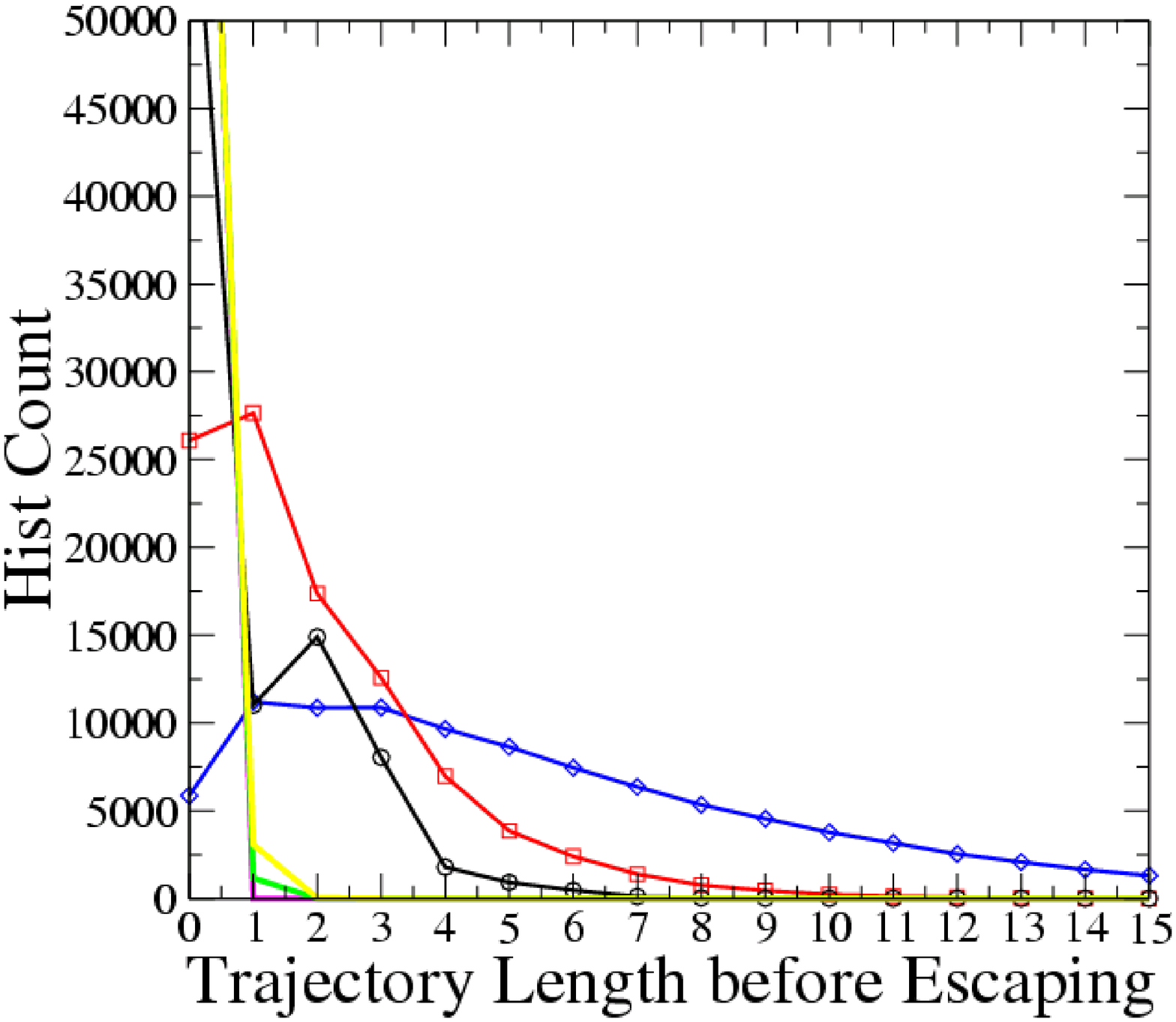}
\hspace{0.in}
\includegraphics[width=5.4cm]{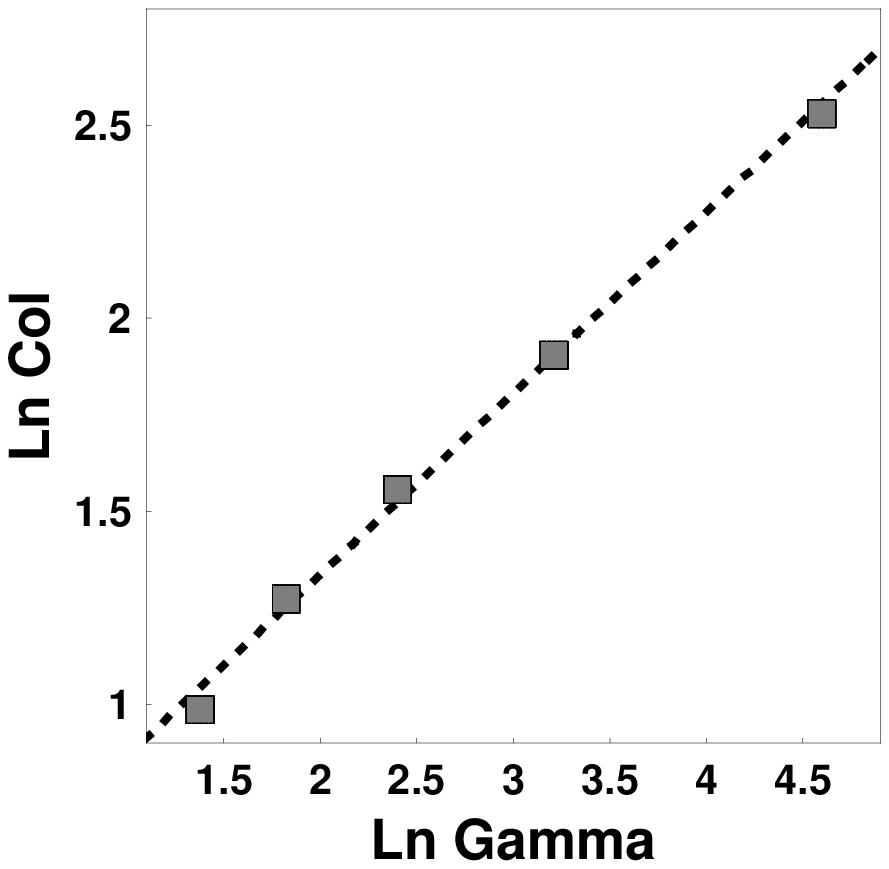}}
\caption{Monopole motion inside a ``cage'' of eight static
electric charges, see text for details.} \label{trapping}
\end{figure*}

In \cite{Liao:2006ry} we found that a mixture plasma of both
electric and magnetic charges has the smallest viscosity and
diffusion (the shortest mean free path, if one wants to use such
language) at the E/M density ratio 50\%-50\% (i.e. maximal
mixing), with the values close to the empirical numbers extracted
from RHIC data. So, what will be the microscopic origin of such
behavior? Due to the E/M ratio dependence, we suggest the Lorentz
force between the two types of particles is the mechanism at work
in similar way to the ``magnetic bottle'' effect originally
invented by G.Budker in 1950's for confining hot plasma. In
\cite{Liao:2006ry} we already showed that in the static electric
dipole field a nearby monopole is ``focused'' by the Lorentz force
to collide head-on with the standing charges and bounce back and
forth.

We here use a ``Gedanken experiment'' to elucidate such effect as
the microscopic origin of the nearly perfect fluidity. We consider
a monopole at the center of a ``grain of salt'' which has eight
static electric charges (with alternating signs) at the corners,
and then we ``kick'' it off with random initial velocity, see
Fig.\ref{trapping}(upper left). This can be repeated many times,
and what we want to learn include: 1) what the trajectory (as
determined by classical equation of motion) will look like; 2) how
long it will typically take for the monopole to escape the cube.

Opposite to naive expectation, it turns out most of the
trajectories are highly complicated: an example is shown in
Fig.\ref{trapping}(upper right). The multiple-folded trajectory
shows nontrivial features: apparently the monopole experiences
many collisions before finally finding the ``door'' out; the
collisions are strong as can be seen from several complete
bouncing back and from its many highly curled parts; also there
are a few clearly visible Poincare-cone like structures near the
corners (where the electric charges are). From this we see that
the monopole, rather than encountering the electric charges at
corners by chance, is focused to rotate on the Poincare-cone all
the way to the charge and then bounced back, only to be focused
toward another corner for the next collision. Such phenomenon is
absent if we replace the monopole with an electric charge. The
Lorentz force here provides a unique way of enhancing the
collision rate and trapping the monopole for long time (not
permanent though): thus we may call it a Lorentz trapping effect.

To quantify the effect we choose different values of initial
velocity magnitude $v_0$, and for each $v_0$ we repeat the
experiment with random initial directions for $10^5$ times and
register for each trial the total trajectory length $L_{Esc.}$
before the monopole escapes the cube. The resulting histograms for
$L_{Esc.}/(2a)$ (with $2a$ the cube side length) are shown in
Fig.\ref{trapping}(lower left) for $v_0=0.1$(blue
diamonds),$0.3$(red boxes),$0.5$(black circles) respectively. The
plots show that $L_{Esc.}/(2a)$, an estimate of collision numbers,
are often much larger than 1. The plots also show that with
smaller $v_0$, $L_{Esc.}$ becomes much flatter, i.e. with more
probability to be trapped for longer time, because monopole with
smaller velocity is more easily curled with smaller Larmor circle
to collide with the charges. For comparison we did the same
experiment for an electric charge replacing the monopole and found
it always exits immediately and never gets bounced back, see the
three indistinguishable curves (yellow, green and magenta) very
close to the left axis. This study demonstrates that Lorentz force
does provide an efficient mechanism that significantly enhances
collision rates and traps particles locally for a time scale
longer than microscopic motion time scale, i.e. $\tau_{Esc.}\equiv
L_{Esc.}/v_{0}>> 2a/v_0 $.

We can define an effective collision number for the monopole with
each given $v_0$ by averaging out the histogram for $L_{Esc.}$:
$C(v_0)\equiv < L_{Esc.}(v_0)/(2a) > $. If we ``pretend'' the
monopole is taken as a representative of a plasma with
$\Gamma=PE/KE$, then $v_0 \propto KE^{1/2} \propto
1/\Gamma^{1/2}$. Thus we obtain a plot showing how $C$ changes
with $\Gamma$, see Fig.\ref{trapping}(lower right). It shows a
linear relation in the Log-Log plot, which is nicely fitted by
$C\propto \Gamma^{0.47}$. The monopole mean free path is then
$L_{MFP}\propto 1/C \propto 1/\Gamma^{0.47}$. One may image that
there is a whole crystal with periodically repeating electric cube
and the monopole is jumping from the original cube to the
neighboring cubes and eventually diffuses away. A hand-waving
argument leads to the diffusion constant for such a monopole:
$D\propto L_{MFP} \propto 1/\Gamma^{0.47}$ which is close to the
power law obtained both from our MD ($1/\Gamma^{0.63}$
\cite{Liao:2006ry}) and from the AdS/CFT calculation
($1/\lambda^{0.5}$ \cite{Teaney}).

Finally consider a dynamical mixture plasma with each charge
surrounded by the charges of the other type, so by the Lorentz
trapping effect both types of charges are spatially interlocked
for much longer time than microscopic motion and diffuse away only
after even longer time. This may be the underlying picture of the
QCD fluid near $T_c$.

\begin{acknowledgments}
This work was supported in part by the US-DOE grant
DE-FG-88ER40388. JL thanks the hospitality of the Institute for
Nuclear Theory and the organizers of the QCD Critical Point
Workshop during which part of this write-up was done. He is also
grateful to M. Baker for interesting discussion.
\end{acknowledgments}

\end{document}